\documentclass[twocolumn,prl,aps]{revtex4}
\usepackage{graphicx}
\begin{document}
\renewcommand{\thefootnote}{\fnsymbol{footnote}}
\sloppy
\newcommand \be{\begin{equation}}
\newcommand \bea{\begin{eqnarray} \nonumber }
\newcommand \ee{\end{equation}}
\newcommand \eea{\end{eqnarray}}
\newcommand{\rar}{\rightarrow}
\newcommand{\eq}{equation}
\newcommand{\eqs}{earthquakes}
\newcommand{\rp}{\right)}
\newcommand{\lp}{\left(}

\title{Is Earthquake Triggering Driven by Small Earthquakes?}
\author{Agn\`es Helmstetter} 
\altaffiliation[Now at ]{Institute of Geophysics and Planetary Physics,\\ 
University of California Los Angeles}
\affiliation{Laboratoire de G{\'e}ophysique Interne et Tectonophysique,
Observatoire de Grenoble, Universit\'e Joseph Fourier, France}

\date{\today}

\begin{abstract}
Using a catalog of seismicity for Southern California, we measure how 
the number of triggered earthquakes increases with the earthquake magnitude.
The trade-off between this relation and the distribution of earthquake 
magnitudes controls the  relative role of small compared to large earthquakes.
We show that seismicity triggering is driven by the smallest 
earthquakes, which trigger fewer events than larger earthquakes, 
but which are much more numerous.
We propose that the non-trivial scaling of the number of triggered 
earthquakes emerges from the fractal spatial distribution of seismicity.
\end{abstract}

\maketitle

\vskip 0.5cm

Large shallow earthquakes are always followed by aftershocks, that are 
due the stress change of the mainshock.
The number $n(M)$ of aftershocks of a mainshock of magnitude $M$ has been
proposed to scale with  $M$ as 
\cite{U69,KK87,K91,R85,R99,SS88,O88,RJ89,YS90,DF91,MD92,HZK00,DL01,F02}
\begin{equation}
 n(M) \sim 10^{\alpha M} ~.
\label{a}
\end{equation}
This relation accounts for the fact that large earthquakes trigger
many more aftershocks than small earthquakes.
A similar relation holds for the distribution of earthquake magnitudes 
$P(M)$ \cite{GR49} given by
\begin{equation}
P(M) \sim 10^{-bM} ~,
\label{GR}
\end{equation}
with  $b$ typically close to $1$, which implies that small earthquakes 
are much more frequent than large earthquakes.

Because large earthquakes release more energy and trigger more aftershocks than
smaller earthquakes, it is usually accepted that interactions between earthquakes
and earthquake triggering are dominated by the largest earthquakes.
However, because they are much more frequent that larger earthquakes, 
small earthquakes are also just as important as large earthquakes in 
redistributing the tectonic forces if $b=1$ \cite{H92}.
Other quantities, such as the Benioff strain $\epsilon \sim 10^{0.75M}$,
are dominated by small earthquakes if $b>0.75$.

The $\alpha$-exponent is an important parameter of earthquake interaction
that is used in many stochastic models of seismicity or prediction
algorithms \cite{KK87,K91,R85,R99,O88,RJ89,CM01,F02}.
This parameter controls the relative role of small compared to large earthquakes.
While there is a significant amount of literature on the $b$-value, very few studies
have measured accurately the $\alpha$ exponent in real seismicity data.
Many studies use $\alpha=b$ without justification 
\cite{KK87,RJ89,DF91,CM01,F02}.
In this case, small earthquakes are just as important as larger ones 
 for the triggering process.
Using (\ref{a}) and (\ref{GR}), the global number $N(M)$ of aftershocks 
triggered by all earthquakes of magnitude $M$ scales as
\begin{equation}
N(M)= P(M) ~ n(M) \sim  10^{(\alpha-b) M} ~,
\label{NM}
\end{equation}
and is indeed independent of $M$ in the case $\alpha=b$.
In the case $\alpha<b$, aftershock triggering is controlled by the
smallest earthquakes, while the largest earthquakes dominate if $\alpha>b$.

A few studies measured directly $\alpha$ from aftershocks sequences,
using a fit of the total number of aftershocks as a function of the mainshock
magnitude \cite{SS88,YS90,MD92,DL01}.
These studies yield $\alpha$-value close to 1, but the limited
range of the mainshock magnitude considered and the large scatter 
of the number of aftershocks per mainshock  do not allow an accurate 
estimation of $\alpha$.
The case $\alpha=b$ also explains another well documented property of
aftershocks, known as Bath's Law \cite{B65,DL01,F02}, which states that
the difference between the mainshock  magnitude and its largest aftershock
is on average close to 1.2, independently of the mainshock magnitude.
Again, the limited range of mainshock magnitudes used in these studies
and possible biases of data selection 
\cite{V69}  does not allow to test the dependence of 
the magnitude difference as a function of the mainshock magnitude.

Other studies measured $\alpha$ indirectly using a stochastic triggering  
model called ``Epidemic Type Aftershock Sequence'' model (ETAS)
\cite{O88,K91,GO97} based only on the Gutenberg-Richter and Omori laws 
\cite{KK87,O88}. This model assumes that each earthquake above a
magnitude threshold $m_0$ can trigger direct aftershocks, with a rate that
increases as  $\sim 10^{\alpha M}$ with its magnitude, and decays with
 time according to Omori law \cite{O94}. The average total number of 
aftershocks $n(M)$, including the cascades of indirect aftershocks, has the
same dependence   $\sim 10^{\alpha M}$  with the mainshock magnitude
$M$ as  the number of direct aftershocks.
Using this model, $\alpha$ can be measured using a maximum likelihood
method  \cite{O88,K91,GO97}.
For instance, \cite{GO97} analyzed 34 aftershock sequences in Japan
and measured $\alpha$ in the range $[0.2-1.9]$ with a mean value of
0.86. The $\alpha$-values obtained from the 
inversion of this model are not well constrained due to the small
number of events available and to possible biases of the inversion method.
Indeed, these studies do not take into account the influence of earthquakes 
below the detection threshold, which may significantly bias the
estimation of $\alpha$. This method may also be biased by the
incompleteness of the catalog just after the mainshock, and by
possible trade-offs between  the ETAS parameters.
The regime $\alpha \geq b$ of the ETAS model is probably not relevant for 
real seismicity. If we do not assume a roll-off of the magnitude distribution
$P(M)$ for large $M$, this regime gives a finite time singularity
of the seismicity rate which goes to infinity in finite time $t_c$
as $1/(t_c-t)^m$ \cite{SH02}. Such a power-law increase of seismic activity 
can describe the acceleration of the deformation preceding 
material failure as well as a starquake sequence \cite{SH02},
but cannot describe a stationary seismic activity.

In this study, we use a stacking method to estimate the average 
rate of earthquakes triggered (directly or indirectly) by a previous
earthquake as a function of the magnitude of the triggering earthquake.
We use the seismicity catalog of Southern California provided by the 
Southern California Data Center \cite{SCEC}, which covers
the time period 1975-2003, and which is complete above $M=3$ for
this time period. The magnitude distribution shown in Fig. 1b
follows the Gutenberg-Richter law for $M \geq 3$ and attests for the 
completeness of the catalog for this time period above magnitude 3.
We do not use the usual distinction between ``foreshocks'',
``mainshocks'' and ``aftershocks'', and the constraint that the
aftershocks must be smaller than the mainshock because this
classification is not based on physical differences.
Indeed, recent studies have shown that a simple model 
that assumes that each earthquake can trigger earthquakes of any
magnitude, without any distinction between ``foreshocks'', ``mainshocks'' 
and ``aftershocks'' can reproduce many properties of real seismicity
including realistic foreshock sequences \cite{KK87,K91,O88,HSominv,F02,HSG03}.
Constraining ``triggered earthquakes'' to be smaller than the mainshock
would obviously underestimate the number of earthquakes triggered
by small mainshocks and thus overestimate $\alpha$.

We define a ``triggered earthquake'' as any event occurring in a 
space-time window $R \times T$ after a preceding ``mainshock'' above the 
background level, whatever the relative magnitude of the triggered and 
triggering earthquakes. We consider as a potential ``mainshock'' each
earthquake that has not been preceded by a previous larger earthquake in a
space-time window  $d \times T$ in order to estimate the rate of 
seismicity triggered by this ``mainshock'' removing the influence
of previous earthquakes.

This definition of ``triggered earthquakes'' and ``mainshocks'' 
contains unavoidably a degree of arbitrariness in the choice of the
space-time windows but the estimation of $\alpha$ is found to be robust
when changing  $T$, $R$ and $d$. We have tested different methods for 
the choice of $R$, either fixed or increasing with the mainshock magnitude.
We  use a distance $R$ increasing with the mainshock magnitude
because the aftershock zone is usually found to scale with the rupture length
\cite{U61,K02}. We use $R$ equal to 1 rupture length of the mainshock.
For small mainshock magnitudes, 
this choice would lead to unacceptable values of $R$ smaller than the 
location error, and thus to  underestimate the number of triggered earthquakes of 
small mainshocks. Therefore, we impose $R>5$ km, larger than the location error.
Taking $R$ fixed has the advantage of not introducing by hand any scaling 
between the aftershock zone and the mainshock magnitude. 
However, it may  overestimate the number of earthquakes triggered by
the smallest mainshocks if $R$ is too large, or underestimate
the number of triggered events of the largest mainshock if $R$ is too small.

The results obtained for $T=1$ year, $R=0.01 \times 10^{0.5 M}$ km and 
$d=50$ km are presented in Fig. 1. The rate of triggered earthquakes is found  
to decay according to Omori's law $K(M)/t^p$, with the same exponent
$p \approx 0.9$ for all mainshock magnitudes $M$ (Fig. 1a). The amplitude 
$K(M)$ increases exponentially $\sim 10^{\alpha M}$ as a function of
$M$ with $\alpha=0.81$  (Fig. 1b). This confirms that the scaling of the 
rate of triggered earthquakes with $M$ follows  (\ref{a}). 
Our method is more accurate than previous studies  \cite{SS88,YS90,MD92,DL01}. 
Indeed, these studies \cite{SS88,YS90,MD92,DL01} determine the scaling of 
$n(M)$ with $M$ using the total number of aftershocks 
\cite{SS88,YS90,MD92,DL01} in a time window $[0-T]$ after a mainshock, 
and can thus be biased by the incompleteness
of the catalogs just after a large mainshock or by the background seismicity 
at large times after a 
mainshock. In contrast, in order to deal with these problems, we use the 
seismicity rate in the time window where we observe the Omori law decay 
characteristic of triggered seismicity.

The value of $\alpha$ is robust when increasing or decreasing the
distance $R$ used for the selection of triggered earthquakes between
1 to 5 rupture lengths, or when increasing the minimum value of $R$
from 2 to 10 km.
Selecting earthquakes within a disk of fixed radius $R=50$ km for 
all mainshock magnitudes yields a slightly smaller value $\alpha=0.72$.
Decreasing $R$ leads to a smaller value of $\alpha$ because it 
underestimates the number of events triggered by the largest mainshocks, 
which  have a rupture size larger than $R$.
When increasing $R$ from 10 to 100 km, the value of $\alpha$ first 
increases with $R$ and then saturates around $\alpha=0.72$ for $R\geq$ 50 km.
We have also checked that $\alpha$ is not sensitive to 
the parameter $d$ used for the selection of mainshocks if $d \geq 50$ km.
All values of $\alpha$, for reasonable values of the parameters 
$R$ in the range $30-100$ km, $T$ between 0.1 and 2 yrs and $d> 50$ km, 
and for different time periods of the catalog, are in the range $[0.7-0.9]$.
We have also tested the method on synthetic catalogs generated with
the ETAS model. We recover the $\alpha$ parameter with an error 
smaller than 0.05.

For the same catalog of seismicity, we measure using a maximum likelihood 
method  the  $b$-value of the 
Gutenberg-Richter law  (\ref{GR}) equal to $b=1.08 \pm 0.10$.
We have also tested that the magnitude distribution $P(M)$ 
of triggered events is independent of the mainshock magnitude.
Fig. 2 shows the magnitude distribution of triggered events for different 
ranges of the mainshock magnitude, using the same data as in Fig. 1.
This figure shows that a large earthquake can be triggered by 
a smaller earthquake.
Our results suggest that $\alpha$ is significantly smaller than 
the $b$ exponent of the magnitude distribution. 
Whether the exponent $\alpha$ varies with region and maybe even with time is an
interesting question that is outside the scope of this Letter
but we urge further studies in that direction.

We now propose a simple explanation for this non-trivial scaling of the 
number of triggered earthquakes with the mainshock magnitude, and we suggest that 
$\alpha$ can be related to the fractal structure of the spatial distribution of 
seismicity. It is widely accepted that the aftershock zone scales with the 
rupture length \cite{U61,K02}. 
While the area affected by the stress variation induced by an earthquake
increases with the rupture length, the stress drop is independent 
of the mainshock magnitude \cite{KA75,IB01}. 
The stress variation at a distance from the mainshock proportional 
to the fault length $L$ is thus independent of the mainshock magnitude,  
neglecting the effect of the
finite width of the crust and the visco-elastic deformation in the lower-crust.
Therefore, assuming that earthquakes triggered by the stress change induced
by the mainshock, the density of earthquakes triggered at a distance 
up to $R \approx L$ from the mainshock is independent of the mainshock magnitude.
The increase of the number of triggered events with the mainshock magnitude results only
from the increase in the aftershock zone size with the rupture length.

The rupture length is usually related to the magnitude by \cite{KA75}
\begin{equation}
L \sim 10^{0.5 M}~.
\label{ace}
\end{equation}
The same relation thus holds between the aftershock zone size $R$ and 
the mainshock magnitude.

In order to estimate the scaling of the number of triggered events with 
the rupture length, we need to make an assumption about the spatial 
distribution of earthquakes around the mainshock.
Assuming that triggered earthquakes are uniformly distributed on the fault plane, 
and using (\ref{ace}), the number of earthquakes triggered by a mainshock 
of magnitude $M$ is given by 
$n(M) \sim L^2 \sim 10^M$ \cite{YS90}
and thus leads to $\alpha=1$.
The value $\alpha=0.5$ obtained for a numerical model of seismicity \cite{HZK00}
suggests that in this model earthquakes are triggered mostly on the edge 
of the fracture area of the mainshock \cite{HZK00}.
Our result $\alpha=0.8$ for Southern California seismicity implies that 
triggered earthquakes are distributed neither uniformly on the rupture 
plane nor on the edge of the rupture, but rather on a fractal structure 
of dimension $D<2$. Using the definition of the capacity fractal dimension, 
the number of aftershocks is
\begin{equation}
n(M) \sim R^D ~,
\label{ode}
\end{equation}
where $R$ is the characteristic length of the aftershock zone.
Using (\ref{ace}) and (\ref{ode}), we obtain the scaling of the number of 
triggered earthquakes with the mainshock magnitude
\begin{equation}
n(M) \sim 10^{0.5 D M}
\label{sfd}
\end{equation}
which gives $\alpha=0.5D$.
Our estimation $\alpha=0.8$ for Southern California seismicity thus suggests $D=1.6$.
This value of the fractal dimension of aftershocks hypocenters has never been
measured for Southern California seismicity.
Our estimate of $D$ is significantly smaller than the value measured in the range
$[2-2.8]$ for aftershock sequences in Japan \cite{GO97}. 
This fractal dimension of the spatial distribution of triggered earthquakes 
results in part from the fractal structure of the fault system \cite{B01},
but it may also reflect the non-uniformity of the distribution of the 
earthquakes on the fault due to the heterogeneity of stress or strength on the fault.
The fractal dimension of the aftershock distribution may thus be smaller than
the fractal distribution of the fault system.

While the energy release and the total slip on faults is controlled by
the largest earthquakes, the suggestion that $\alpha<b$ implies that small 
earthquakes may be more important than large earthquakes in triggering 
earthquakes. We have also checked that the magnitude distribution of triggered 
earthquakes is independent of the mainshock magnitude (Fig. 2).
This implies that earthquake triggering is driven by the smallest 
earthquakes {\it at all scales}, even for the largest earthquakes. 
Other observations  \cite{HSominv} support the conclusion 
that the same mechanisms can explain the triggering of a large earthquake 
by a smaller one and the triggering of a small 
earthquake  by a previous larger event.

Recent studies \cite{F02} have proposed that secondary aftershocks 
dominate an aftershock sequence, so that subsequent large aftershocks are 
more likely to be triggered indirectly by a previous aftershock of the mainshock.
Our study further suggests that the smallest earthquakes will dominate 
the triggering of following earthquakes. 
The importance of small earthquakes  casts doubts 
on the relevance of calculations of direct stress transfer functions 
to predict seismicity \cite{stein}, because large earthquakes 
are likely to be triggered by the smallest earthquakes
below the detection threshold of the seismic network.
Small earthquakes taken individually have a very low probability of triggering
a large earthquake. But because they are much more numerous than larger earthquakes,
collectively, they trigger more earthquakes.
This result requires the existence of a small magnitude cut-off $m_0$, 
 below which earthquakes may occur but cannot trigger earthquakes larger than $m_0$, 
or a change of the scaling of $N(M)$ given by (\ref{NM}) for small earthquakes,
otherwise the seismicity at all scales would be controlled by infinitely 
small earthquakes.

 I am very grateful to D. Sornette, J.-R. Grasso, A. McGarr, L. Margerin, 
C. Voisin and T. Gilbert for useful suggestions and discussions.
I am also grateful for the earthquake catalog data made available 
by the Southern California Earthquake Data Center.

{}
\begin{figure}
\includegraphics[width=16cm]{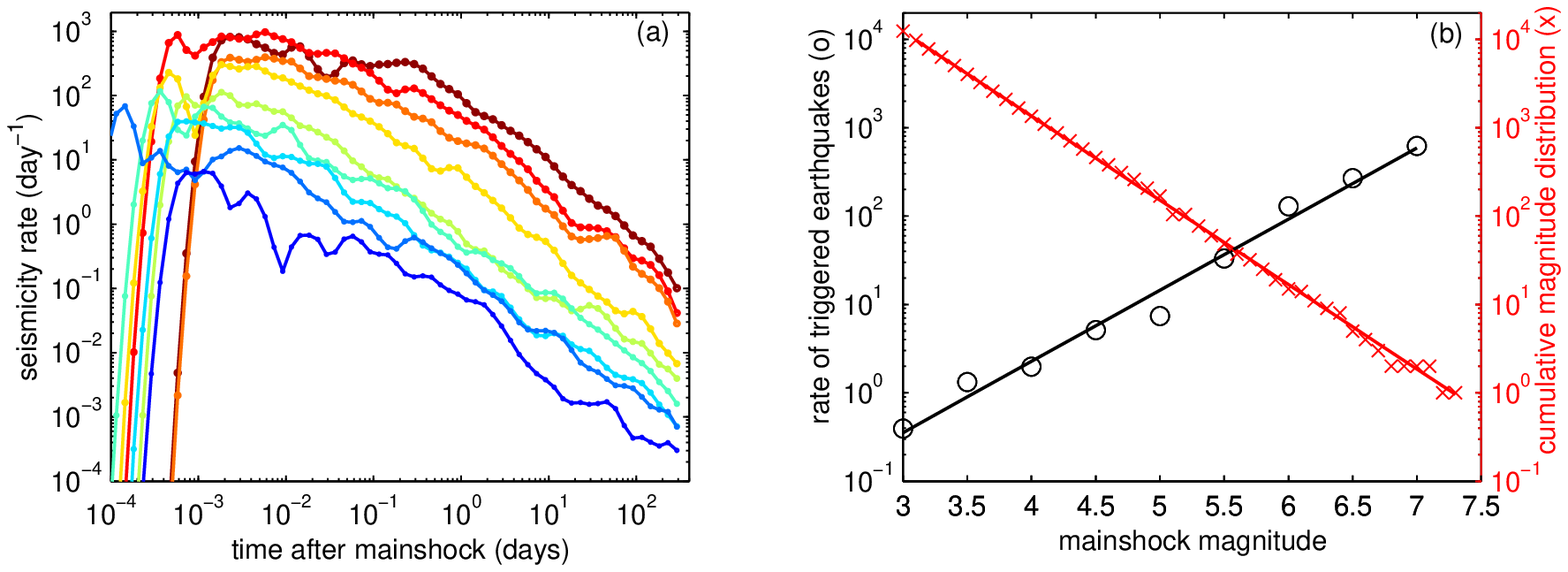}
 \caption{Average rate of triggered earthquakes $n(M,t)$
as a function of time $t$ after the triggering earthquake (a)
for different values of the magnitude $M$ of the triggering earthquake
 increasing from 3 to 7 with a step of 0.5 from bottom to top.
The rate of aftershocks $K(M)$ as a function of $M$ is shown in panel
(b) (circles) with the cumulative magnitude distribution (crosses).
$K(M)$ is obtained by fitting each curve  $n(M,t)$ by $K(M)/t^{0.9}$
in the range $0.01<t<365$ days for $M<6.5$.
For large $M \geq 6.5$ mainshocks, there is a roll-off of the seismicity rate
for small times after the mainshock due to the incompleteness of the
catalog after large mainshocks, caused by the saturation of the
seismic network. Therefore we measure $K(M)$ in the range $t>0.1$ day
for $M=6.5$ and $t>0.3$ day for $M=7$.}
\label{fig1}
\end{figure}

\begin{figure}
\includegraphics[width=8cm]{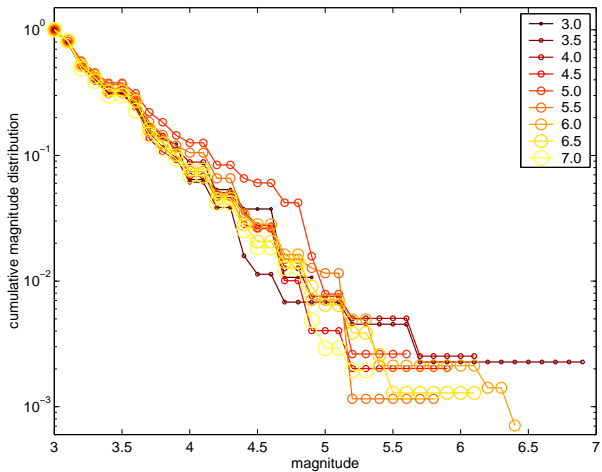}
\caption{Cumulative magnitude distribution of triggered earthquakes 
for different values of the mainshock magnitude between 3 
(dark line, small circles) and 7 (gray line, large symbols)
using the same time interval for the selection of aftershocks 
as  in Fig. \ref{fig1}.}
\label{fig2}
\end{figure}

\end{document}